\documentstyle[12pt,aasms4]{article}

\def\gtsim{\ {\raise-0.5ex\hbox{$\buildrel>\over\sim$}}\ }
\def\ltsim{\ {\raise-0.5ex\hbox{$\buildrel<\over\sim$}}\ }

\begin{document}

\title{A New Giant Branch Clump Structure In the Large Magellanic Cloud}

\author{Andr\'es E. Piatti\altaffilmark{1,2} and Doug Geisler}
\affil{Universidad de Concepci\'on, Departamento de F\'{\i}sica, 
Casilla 160-C, Concepci\'on, Chile;\\
piatti@gemini.cfm.udec.cl,doug@stars.cfm.udec.cl}

\author{Eduardo Bica}
\affil{Departamento de Astronomia, Instituto de F\'{\i}sica, UFRGS,
        C.P. 15051, 91501-970  \\
Porto Alegre, RS, Brazil;\\
bica@if.ufrgs.br}

\author{Juan J. Clari\'a\altaffilmark{1}}
\affil{Observatorio Astron\'omico de C\'ordoba, Laprida 854, 5000, C\'ordoba,
Argentina;\\
claria@oac.uncor.edu}

\author{Jo\~ao  F. C. Santos Jr.\altaffilmark{1}}
\affil{Dep. de F\'{\i}sica, ICEx, UFMG,
C.P. 702, 30123-970 Belo Horizonte, MG, Brazil;\\
jsantos@fisica.ufmg.br}

\author{Ata Sarajedini}
\affil{Astronomy Department, Van Vleck Observatory, Wesleyan University, \\
Middletown, CT 06459;\\
ata@urania.astro.wesleyan.edu}

\and

\author{Horacio Dottori}
\affil{Departamento de Astronomia, Instituto de F\'{\i}sica, UFRGS,
        C.P. 15051, 91501-970  \\
Porto Alegre, RS, Brazil;\\
dottori@if.ufrgs.br}

\slugcomment{To be submitted to the Astronomical Journal}

\altaffiltext{1}{ Visiting Astronomer, Cerro Tololo Inter-American
Observatory, which is operated by AURA, Inc., under cooperative
agreement with the NSF.}

\altaffiltext{2}{Gemini Fellow, NOAO/CTIO-Universidad de Concepci\'on, Chile.}
\begin{abstract}

\end{abstract}

We present Washington $C,T_1$ CCD photometry of 21 fields located in the
northern part of the Large Magellanic Cloud (LMC), and spread over
a region of more than 2.5$^{{\Box}^{o}}$ approximately 6$^o$ from the bar.
The surveyed areas were chosen on the basis of their proximity to SL\,388 and
SL\,509, whose fields showed the presence of a secondary 
giant clump, observationally detected by Bica et al. (1998, \aj, 116, 723). We 
also observed NGC\,2209, located $\sim$14$^o$ away from SL\,509.
>From the collected data we found that most of the observed field CMDs do not 
show a separate secondary clump, but rather a continuous vertical  structure (VS), 
which is clearly seen for the first time. The VS 
also appears in the field of NGC\,2209. Its position and size  are
nearly the same  throughout the surveyed regions: it lies below the Red
Giant Clump (RGC) and extends from the bottom of
the RGC to $\sim$ 0.45 mag fainter, spanning the bluest color range of
the RGC. In two fields as well as in the NGC\,2209
field, the RGC is slightly tilted, following approximately the reddening
vector, while the VS maintains its verticality. We found that
the number of stars in the VS box defined by $\Delta(C-T_1)$ = 1.45-1.55 mag and
$\Delta T_1$ = 18.75-19.15 mag has a strong spatial variation,
reaching the highest VS star density just north-east of SL\,509.
Moreover, the more numerous the VS stars in a field, the larger the number of LMC
giants in the same zone. We also found that, in addition to SL\,509, two relatively 
massive star clusters, SL\,515 and NGC\,2209, separated by more than ten degrees 
from each other, develop giant clumps with a considerable number 
of VS stars. This result demonstrates that VS stars belong to the LMC 
and are most likely the result of some kind of evolutionary process in the LMC,
particularly in those LMC regions with a noticeable large giant
population. Our results are successfully predicted by the models of Girardi
(1999, MNRAS, submitted) in the sense that a large proportion of 1-2 Gyr old stars 
mixed with older stars, and with metallicities higher than [Fe/H] $\simeq$ -0.7 
should result in a fainter and bluer secondary clump near the mass where
degenerate core He burning takes place. However, our results apparently 
suggest  that in order to trigger the formation of VS stars, there should be other conditions besides 
the appropriate age, metallicity, and the necessary red giant star density. 
Indeed, stars satisfying the requisites mentioned above are commonly found 
throughout the LMC, but the VS phenomenon is only clearly seen in some 
isolated regions. Finally, the fact that clump stars have an intrinsic luminosity dispersion
further constrains the use of the clump magnitude as a reliable distance indicator.

keywords : galaxies: indivudual (Magellanic Cluds) --- galaxies: photometry ---
           galaxies: stellar content

\eject

\section{INTRODUCTION}

The Large Magellanic Cloud (LMC) has long been a favorite stellar laboratory,
providing us not only with valuable information about its own complex
star formation history but also with important clues for understanding
the formation and evolution of distant galaxies. Moreover, the 
interest in studying different astrophysical aspects of this galaxy has been 
rapidly increasing recently, mainly due to the advent of more 
powerful telescope/instrument combinations and computing facilities.

Recently, Geisler et al. (1997, hereinafter Paper\,I) carried out a search for the  
oldest star clusters in the LMC by observing  with the Washington $C,T_1$ 
filters candidate old clusters spread  throughout the LMC disk. Although
they did not find any genuine old cluster like the Galactic globular clusters,
their study has considerably increased the sample of intermediate-age clusters
($t\sim$ 1-3 Gyr) with ages determined with a high degree of confidence. In 
addition, their results reinforce the conclusion that an important epoch 
of cluster formation, which began $\sim$ 3 Gyr ago, must have been preceded 
by a quiescent period of many billion years, unless dissipation processes have 
been more  effective than previously thought (e.g., Olszewski 1993). In addition, 
they determined not only the properties of the clusters 
but also those of their surrounding fields. From the relatively 
wide field covered by their images ($\sim$ 15$\arcmin$ on a side) they found
that clusters and fields have on average similar ages and metallicities,
except in 3 cases where clusters are $\approx$ 0.3 dex more metal-poor
than the surrounding field, suggesting that the  chemical evolution was not globally
homogeneous (Bica et al. 1998, hereinafter Paper\,II).

A further intriguing result of Paper\,II was the discovery of 
what appeared to be a well populated secondary clump in the Color-Magnitude 
Diagrams (CMDs) of two fields located in  the northern part of the LMC near 
the clusters SL\,388 and SL\,509. This unusual feature, made up by stars 
distributed uniformly across the fields, lies below the prominent Red Giant 
Clump (RGC) slightly toward its bluest color and extends 0.45  mag fainter. 
The feature also appears in the very populous SL\,769 field located $\sim$ 
6$^o$ away, thus representing  around 10\% of the whole sample of fields 
observed  in Paper\,I. Since this feature appeared as a roughly distinct 
secondary clump, Bica et al. coined the term ``dual clump'' to describe this
phenomenon. The authors tentatively suggested that these stars are 
evidence of a depth effect with a secondary component located behind the LMC 
disk at a distance comparable to the Small Magellanic Cloud (SMC), perhaps due to 
debris from previous interactions of the LMC with the Galaxy and/or the SMC.
However, they also mentioned arguments against this scenario and noted other 
possible explanations.

Westerlund et al. (1998) have also found a similar feature
in the CMDs of  three fields located in the NE of the LMC. On the basis of
their $BV$  photometry they suggested that the red giant clump
is bimodal and contains stars from an old population ($t$ $\ge$ 10 Gyr)
and from another younger population ($t$ $\ge$ 0.3-4 Gyr), in the sense that
the fainter the clump the older the stars. Besides observational
findings, Girardi et al. (1998) and Girardi (1999) have theoretically 
predicted
that stars slightly heavier than the maximum mass for developing degenerate
He cores should define a secondary clumpy structure, about 0.3-0.4 mag in
the $I$ band below the bluest extremity of the red clump. According to
Girardi (1999) this evolutionary effect should be seen in CMDs of composite
stellar populations containing $\sim$ 1 Gyr old stars and with mean
metallicities higher than $Z$ = 0.004. However, the current state of both
observational and theoretical  results makes it impossible to determine
whether the intriguing feature is  caused by the presence of an old stellar 
population or by an evolutionary effect or even by a layer of stars located 
behind the LMC. Furthermore, not only its origin but also its morphology
remains uncertain, which must be known before the 
magnitude of the red giant clump can be used as a robust distance indicator 
(e.g., Paczy\'nski \& Stanek 1998).

In this paper we report on the first observations carried out 
with  the aim  of mapping  the extent and determining the nature of the 
``dual clump'' phenomenon. Indeed, the apparent dual clumps
from the limited sample of Paper\,II are now found to merge and form a
continuous feature. The selection and observation of the fields, 
as well as the 
reduction of the data  are presented in Section 2. In Section 3 we present the 
results and discuss them in the light of recent theoretical and 
observational interpretations. Finally, in Section 4 we summarize our main 
conclusions.

\section{OBSERVATIONS AND REDUCTIONS}

The fields for mapping out the extent of the secondary clump phenomenon were 
selected on the basis of their proximity to SL\,388 
and SL\,509, and the presence of star clusters which had not been observed with
Washington photometry. The first criterion aims at observing
LMC regions located not only in the line of sight between SL\,388 and SL\,509,
but also those placed around them within  one degree from the midpoint of 
both clusters. The nearest cluster from this center from Paper\,II is SL\,262 
which is located  1.5$^o$ from SL\,388 and no dual clump structure is visible 
in its CMD. In order to maximize the assigned telescope time, we centered the 
fields so that they included  clusters  without, or with only very unreliable, age 
and  metallicity determinations for our continuing study of the chemical evolution
of the LMC. The clusters having integrated $UBV$ photometry were taken from Bica 
et al. (1996), and the fainter ones from the recent revised catalog of star 
clusters, associations and emission nebulae (Bica et al. 1999).
We also observed the field of NGC\,2209 for the purpose of checking the
possible evolutionary origin of the dual clump. This cluster is located 
$\sim$ 14$^o$ toward the south-east from SL\,509 and placed by Corsi et al. (1994)
photometric data in the
minimum of the relationship between red giant clump and Main Sequence (MS) 
termination magnitudes. Table 1 lists the selected fields and the clusters
contained within these fields. Note that fields \#
5, 16, 17, 23, and 26 were not observed.

The observations were carried out at the CTIO 0.9m telescope during 6
photometric nights in November 1998. The Cassegrain Focus IMager (CFIM)
and the CCD Tek\,2K \#3 were employed in combination with the Washington $C$
and Kron-Cousins $R$ filters. Geisler (1996) has shown that 
the $R_{KC}$ filter is a very efficient and accurate substitute for the 
Washington $T_1$ filter. The pixel size of the detector was 
0.4$\arcsec$/pixel, resulting in a field of $\sim$ 13.5$\arcmin$ wide.
We used the Arcon 3.3 data adquisition system in $quad$ mode (four 
amplifiers) with a mean gain and readout noise of 1.5 $e^-$/ADU and 
4.2$e^-$, respectively. During each night exposures of 2400$^{s}$ in $C$ 
and 900$^{s}$ in $R_{KC}$ were taken for the selected fields as well as for
standard fields (Geisler 1996) with airmasses approximately ranging from 1.1 up to 
1.6. In addition, a series of 10 bias and 5 dome and sky flatfield exposures 
per filter were obtained nightly. The weather conditions kept very stable with a
typical seeing of 1.0$\arcsec$-1.2$\arcsec$, although some images have 
slightly larger FWHMs due to temperature changes of up to 2$^o$ C. In 
general, the secondary mirror was focused twice per night. We covered a total 
area of $\approx$ 1$^{{\Box}^{o}}$ spread over $\sim$ 2.6$^{{\Box}^{o}}$.

The collected data for a total of 21 selected fields and the NGC\,2209 field
were fully processed at the 
telescope using the QUADPROC package in IRAF\footnote{IRAF is distributed 
by the  National Optical Astronomy Observatories, which is operated by the 
Association of Universities for Research in Astronony, Inc., under cooperative
agreement with the National Science Foundation.}. The distribution of the 
observed fields is shown in Fig. 1.  After applying the  
overscan-bias subtraction for the four amplifiers independently, we carried 
out flatfield corrections using a combined skyflat frame, which was
previously checked for non-uniform illumination pattern with the averaged 
domeflat frame. Then, we did aperture photometry for the standard fields 
($\sim$ 30 stars per night) using the PHOT task within DAOPHOT\,II (Stetson 1991).
The relationships between instrumental and standard magnitudes
were obtained by fitting the following equations :

\begin{center}
\begin{equation}
c = a_1 + C + a_2 \times X_C + a_3 \times (C-T_1)
\end{equation}
\begin{equation}
r = b_1 + T_1 + b_2 \times X_R + b_3 \times (C-T_1),
\end{equation}
\end{center} 

\noindent in which $a_i$ and $b_i$ ($i$ = 1, 2 and 3) are the coefficients
derived through the FITPARAM routine in IRAF, and $X$ represents the effective 
airmass. Capital and lowercase letters represent standard and instrumental
magnitudes. The resulting coefficients and their standard deviations are 
listed in Table 2, the typical rms errors of eqs. (1) and (2) being
0.017 and 0.015 mag, respectively. 

Point Spread Function (PSF) photometry for the LMC fields and the NGC\,2209 field 
was performed using the 
stand-alone version of the DAOPHOT\,II package (Stetson 1994),
which provided us with $X$ and $Y$ coordinates and instrumental $c$ and $r$
magnitudes for all the stars identified in each field. The PSFs were generated
from two samples of 30-35 and $\sim$ 100 stars selected interactively. For each 
frame a quadratically varying PSF was derived  by fitting the stars in the larger 
sample, 
once their neighbors were eliminated using a preliminary PSF obtained from the 
smaller star sample, which contained the brightest, least contaminated stars.
Then, we used  the ALLSTAR program for applying the resulting  
PSF to the identified stellar objects and creating a subtracted image which was used 
for finding and measuring magnitudes of additional fainter stars. The PSF magnitudes 
were determined using as zero points the aperture magnitudes yielded by PHOT. 
This procedure was iterated three times on each frame. Next, we computed aperture 
corrections from the comparison of PSF and aperture magnitudes using the subtracted 
neighbors PSF star sample, resulting in typical values around -0.016$\pm$0.010 mag.
Notice that PSF stars are distributed throughout the whole CCD frame, so that
variations of the aperture correction should be negligible. Finally, the standard 
magnitudes and colors for all the measured stars were computed 
inverting eqs. (1) and (2), once positions and  instrumental $c$ and $r$ magnitudes 
of stars in the same field were matched using Stetson's DAOMATCH and DAOMASTER 
programs. Thus, we achieved accurate photometry for $\sim$ 242,000 stars, with
mean magnitude and color errors for stars brighter than V = 19 of 
$\sigma$$(T_1)$ = 0.014 and $\sigma$$(C-T_1)$ = 0.022 mag, respectively.

Later on, with the aim of gathering both astrometric and photometric information in a
self-consistent way, we built a master file which contains the positions for all 
the stars referred to a unique coordinate system. For some fields we only applied
the appropriate offsets in the $X$ and $Y$ values, while in other fields we matched 
from tens up to hundreds of stars in common using DAOMASTER and our own routines. 
We also averaged their $T_1$ magnitudes and $C-T_1$ colors and recomputed their
photometric errors based on the difference. The typical mean difference 
(absolute value) for approximately five hundred stars brighter than V = 19 
in common  turned out to be $\Delta$$T_1$ = 0.026$\pm$0.021 and 
$\Delta$$(C-T_1)$ = 0.030$\pm$0.023. In total, 7760 stars have two
measurements of their magnitude and color. This photometry can be obtained
from the first author upon request. 

\section{ANALYSIS AND DISCUSSION}

\subsection{Description of CMDs}

The CMDs of the 21 observed LMC fields certainly show a mixture of different 
stellar populations. They appear to be dominated by 
a 3-4 Gyr old population as deduced from the $\delta$$T_1$ index, which 
measures the difference in magnitude between the mean magnitude of the 
clump/HB and the MS turnoff 
(see Paper\,I). Fig. 2 illustrates a typical field CMD of the surveyed region. 
Likewise the MS is well-populated and extends along $\sim$ 3 mag.
Assuming that the MS comes from the superposition of MSs with different 
turnoffs, we estimated an age range
from 3 up to 7.5 Gyrs using the $\delta$$T_1$ index, with an average of 
4.5$\pm$1.0 Gyr for all the fields. 
This significant 
age range is also supported by the presence of a Sub Giant Branch 
(SGB) with a broad vertical extension due to the transition of MS stars with
different ages to the SGB. The Red Giant Branch (RGB) is also clearly visible 
covering a wide range in color from $C-T_1$ $\sim$ 1.8 up to 3.6. However, 
the most striking feature of these CMDs is the giant clump region. In addition
to the normal Red Giant Clump (RGC), most of 
the field CMDs also show a vertical structure (VS) composed of stars which lie
below the RGC and extend from the bottom of the RGC to $\sim$ 0.45 mag fainter.
The VS spans the bluest color range of the RGC and also appears in the CMD of
NGC\,2209. This intriguing feature does not clearly appear in the CMDs of our previous LMC 
clusters survey, but only a dual clumpy structure in around  10\% of the cluster
sample (see Paper II). To our knowledge, such a feature has not been observed
previously.

In order to delimit and characterize this intriguing feature, we first estimated
its position and size, and examined its shape going through the individual field 
CMDs. An enlargement of the area of interest in these CMDs is 
shown in Fig. 3. As can be seen, the RGCs are nearly located at the same magnitude 
and color, centered at $\approx$ 18.5 and 1.60 mag, respectively. The 
constancy of the location in the CMD also appears to be the case for the VS, even 
in those fields where the VS arises as a small and sparse groups of stars. In two 
fields  (marked with an asterisk in Fig. 1) the RGC is slightly tilted, following 
approximately the reddening vector, but the mean positions of both the RGC and VS
remain unchanged. Moreover, the VS maintains not only its mean position but also 
its verticality. Therefore, given that the locus of the VS in the CMD does not seem 
to show any correlation with position in the LMC and that reddening variations over 
our survey field should be minimal and, in order to 
highlight the VS phenomenon, we built a composite CMD using all the measured stars.
The resulting diagram is shown in Fig. 4 in which we also included our published
Washington photometry for SL\,338 and SL\,509 (see Section 3.2). We define VS 
stars as those stars which fall into the rectangle $T_1$ $=$ 18.75-19.15 and 
$C-T_1$ $=$ 1.45-1.55. This definition results in a compromise between maximizing 
the number of VS stars and minimizing contamination from, among other sources, MS, 
SGB, RGB, and Red Horizontal Branch stars. The continuous nature of this 
feature is clearly evident.
We are unsure of the reason why this feature appeared as a ``dual clump'' in
two of our Paper\,II fields. In none of our present fields is there any
significant bifurcation. The composite CMD of Fig. 4 thus should present the best
representation of this feature.

\subsection{The VS phenomenon}

It was mentioned above that VS stars appear to be present in most of the fields of 
Fig. 1, although in some of them they could  hardly be recognized. Therefore, it 
would be appropriate to map out the extent of the VS phenomenon in order to have
a more quantitative estimate of its dimensions. For that purpose, we counted 
the number of stars  lying within the VS rectangle, assuming for all the fields 
the same Galactic field star distribution. This assumption is particularly true if the
CMDs of cluster fields located in different parts of the LMC disk (see Fig. 4 in
Paper\,II) are compared with that of the outermost field (OHSC\,37), for which 
we found no evidence of LMC field stars (see Santos et al. 1999, hereinafter 
Paper\,III). Furthermore,  in an area of the same size as the selected LMC fields, 
the OHSC\,37 field  only has two stars within the VS
box, so that we did not perform any correction due to foreground star 
contamination. The number of stars we counted for the northern LMC fields are shown 
in Fig. 1. The VS stars show a {\bf strong spatial variation}, reaching their 
highest density just north-east of SL\,509.

We also repeated the same counting procedure for the fields of SL\,126, SL\,262, 
SL\,388, SL\,509 and SL\,842 by revisiting our Washington data published in 
Paper\,II. SL\,126, SL\,262 and SL\,842 are the nearest clusters to SL\,388 and 
SL\,509. Both present and published data sets were obtained following the same 
stellar object selection criteria, so that they can be compared directly. The number 
of VS stars in SL\,388 and SL\,509 is 37 and 106, which successfully match the trend
followed by the selected LMC fields (see Fig. 1). On the other hand, the fields of 
SL\,126, SL\,262 and SL\,842  turned out to have 4, 9 and 9 VS stars, respectively, 
placing an upper limit to the size of the VS region. All these results 
suggest that the region in which the VS phenomenon is concentrated 
extends over at least $\sim$ 2$^{{\Box}^{o}}$ and that 
the feature is not clearly seen  either in the field of  SL\,262, which is $\sim$ 1.5$^o$ to 
the NW of SL\,388, nor in SL\,842 located $\sim$ 4.5$^o$ to the NE of SL\,509.

With the aim of looking into whether VS stars are also found in LMC
{\bf clusters}, we took advantage of the fact that there is roughly one star cluster
in each selected field; the total cluster sample being 21. First, we
determined the cluster centers and selected their radii by eye-judging the variation
of the stellar density in the cluster surroundings. Cluster radii vary between
20 and 110 pixels with an average of 45 pixels (20$\arcsec$). Then, we performed
VS star counts within both cluster radii and  four  circular LMC field
areas on the same image chosen for comparison purposes. The four circular field areas were distributed 
throughout the entire image; none of them  closer to the corresponding
cluster  than three cluster radii, thus avoiding cluster star contamination. They
were also located far away from the image edges, avoiding vignetting and
flatfield residuals effects. Finally, the radii of the four comparison areas in each
image were fixed at one-half of the cluster radius. Eighty percent
of the selected LMC comparison fields contained no VS stars, while one and two VS stars
were found in two and one LMC comparison fields, respectively. These results are in very good
agreement with the total number of VS stars found in each frame once they are 
scaled to the cluster area. Similarly, star clusters apparently also  have no 
VS stars, except in the case of SL\,515 which has 12 VS
stars, five of them lying inside 1/2 of the cluster radius ($r$$\sim$45$\arcsec$). 
SL\,515 is located in one of the selected LMC fields with the highest VS star
densities as can be seen in Fig. 1. However, 9$\arcmin$ to the SE from 
SL\,515 there is another star cluster (SL\,529) which only has two VS stars. 
Therefore, the excess of VS stars in SL\,515 would not seem to be related to the 
peak in the field VS star distribution, but with some property of the cluster itself,
presumably its mass (see Section 3.3). 
By looking at the images and comparing the cluster radii we 
found that SL\,515 is the largest and perhaps the most massive 
cluster in the sample. We also made the same comparison for the SL\,388 and SL\,509 
fields and found one and five cluster VS stars, respectively, and no field VS stars.
Both clusters have relatively large radii ($r$$\sim$30$\arcsec$).

An additional test for exploring the nature of the VS phenomenon consists of 
comparing the number of VS stars with the total number of stars in the CMDs 
of different regions to investigate whether there is any trend 
of the ratio  between them with their spatial distributions
(No. VS stars/Total number CMD stars $= F$(position)). For this test, we 
decided not to use the whole CMD of each selected LMC field because of different 
incompleteness  factors at fainter magnitudes. Thus we did not consider 
MS stars but  rather a box defined by $T_1$ = 17.5-19.7 and $C-T_1$ = 1.0-2.2,
which are precisely the limits of the CMD in Fig. 4. This box (hereinafter RG box) 
includes all the red giant phases so that if there were 
any correlation between VS and LMC giant stars (strictly VS $= f$(RG-VS)), it should 
arise without any bias 
due to the presence of MS or other kinds of stars. Fig. 5 shows the resulting 
relationship in which we also include the fields of SL\,126, SL\,262 and 
SL\,842.  There is a strong correlation between the number of VS stars in the
field and the number of LMC giants in the same zone. The lowest VS star 
counts occur in the outermost LMC fields, such that of OHSC\,37, where the
number of red giants is also a minimum.

\subsection{The NGC\,2209 case}

The giant clump luminosity is one of the best indicators of the development
of the RGB, and consequently, an important tool for studying the nature of the
VS phenomenon. Indeed, the RGC luminosity varies along a sequence which depends
on the age (mass) of the giant stars. Furthermore,  Corsi et al. 
(1994) data and Girardi (1999) models show that the clump magnitude ($V_{clump}$) 
has a maximum- (faintest value) as a function of the termination MS magnitude 
($V_{TAMS}$) which 
corresponds to an age of $\approx$ 1.0-1.5 Gyr. Precisely, our interest in
observing NGC\,2209 comes from the fact that this cluster falls onto the faintest 
magnitude in the $V_{clump}$ vs. $V_{TAMS}$ relationship shown in Corsi et al.
(1994), thus providing us with a valuable opportunity to test whether the VS is 
caused by evolutionary effects. NGC\,2209 is located $\sim$ 14$^o$ away from the 
selected LMC fields and therefore any local effects in our VS area should
be negligible.

Using our Washington photometry we performed an analysis similar to that 
carried out for the  selected LMC fields, 
i.e., we first looked at the NGC\,2209 field CMD. Its main features
resemble those of the northern LMC fields, as shown in Fig. 6. The RGC is tilted
and shifted with respect to Fig. 1 by $\Delta$$(C-T_1)$ $\approx$ 0.20 and 
$\Delta$$T_1$ $\approx$ 0.30 mag. According to the relations $E(C-T_1)$ = 
1.966$E(B-V)$
and $A_{T_1}$ = 2.62$E(B-V)$  (Geisler 1996) these offsets are consistent with a mean 
reddening $\approx$ 0.10 mag higher. Fig. 6 also reveals the 
presence of a VS at the same position relative to the RGC, reinforcing 
the conclusion that VS stars belong to the LMC. Its shape and magnitude extent are 
essentially the same as described in Section 3.1,  while its color range is somewhat
wider. The tilted RGC following approximately the reddening vector (see Fig. 6)  
could suggest the existence of differential reddening, although evolutionary effects 
could also yield an inclined clump. According to Catelan \& Freitas Pacheco (1996) 
horizontal branch (HB) stars could result in a tilted clump if the helium content 
were very high (Y=0.30). They also argued that a differential reddening 
as small as $\delta$$E(B-V)$ = 0.06 mag cannot cause a CMD dispersion as large as 
the one originating from the evolution away from the Zero Age HB itself. Notice also
that tilted clumps also appear in two fields marked with an asterisk in Fig. 1, but 
their positions and sizes (magnitude and color dispersions) are nearly the same as
the remaining fields (see Section 3.1). On the other hand, Hodge (1960)
noticed an apparently dark patch in NGC\,2209 of $\sim$ 15$\arcsec$ in diameter, about 
10$\arcsec$ from
the cluster center, suggesting either an internal or foreground origin for the
globule. In addition, using $BV$ CCD photometry and CMD analysis, Dottori et al. 
(1987) concluded that the globule should be internal to the cluster, so that
differential reddening is not unexpected. Indeed,  we estimated a VS width 
approximately twice that of the northern selected fields. The extracted CMD of
NGC\,2209 also shows a remarkable color dispersion not only for giant clump stars,
but also for SGB stars, which appear distributed  at both edges of their whole 
color range (see Fig. 6). 

Next, we counted the VS stars distributed in the NGC\,2209 field using a box with the
same dimensions as for the northern LMC fields and reddened by $\Delta$$E(B-V)$ = 
0.10. We also applied the same shift to the RG box, thus centering the RGC.
The number of VS stars in the NGC\,2209 field and in the cluster itself 
($r$$\sim$45$\arcsec$) was 69 and 10, respectively, whereas no VS 
stars were found  in four circular field areas (equal cluster area criterion), as
expected. This result is in very good agreement with that found for SL\,515, 
in the sense that relatively massive clusters can develop giant clumps with a 
considerable number of VS stars. Finally, if we compare the field VS stars number 
with that corresponding to the RG box, we can conclude that LMC regions with a 
noticeable large giant population can also be reservoirs of VS stars. The fact
that NGC\,2209, located many degrees from our main VS area, also shows this
feature argues against a depth effect interpretation (e.g., background 
galaxies or debris) and for an evolutionary origin.

\subsection{Comparison with theory}

It is known that stars defining the RGC in CMDs of intermediate-age 
and old open clusters are in the stage of central 
helium burning (Cannon 1970, Faulkner \& Cannon 1973). However, according to 
Girardi (1999) models - computed using a grid of masses with a resolution of 
$\sim$ 0.1 $M_{\odot}$ in the vicinity of the onset of helium burning mass - the 
position of 
GC stars in the CMD depends on the masses of the stars. Particularly, 
stars with $M \le M_{Hef} \sim 2-2.5 M_{\odot}$ form
electron-degenerate cores with masses nearly constant ($M_c \simeq 0.45 M_{\odot}$)
after the central hydrogen exhaustion, thus allowing stars to reach similar 
luminosities. These stars correspond to our RGC stars. On the other hand, 
for stars with $M >  M_{Hef}$ helium ignition  takes place under 
non-degenerate conditions and both
$M_c$ and luminosity increase with $M_{Hef}$, the minimum luminosity being
about 0.4 mag fainter than those of stars with slightly lower masses. 
Girardi's models predict that such stars should define a secondary clumpy feature 
located below the RGC and at its bluest extremity, reminiscent of our VS feature. 
The spread in the intrinsic luminosity of stars burning helium
in their cores evidenced by this feature provides a further constraint
on using the magnitude of the GC stars as a
self-consistent distance indicator.

Now, we can check Girardi's (1999) predictions in the light of the present 
observational findings, so that new constraints to the theory can 
improve our knowledge of stellar evolution and the star formation history in 
the LMC. In contrast with the tentative explanation of debris from a dwarf galaxy 
located behind the LMC suggested in Paper\,II, Girardi claimed that 
the secondary clump in the CMDs of SL\,388 and SL\,509 fields might have been 
caused by a population younger (higher mass) than RGC stars.  However, in the present 
work we did not find
such a separated fainter clump but rather a VS having approximately the same number
of stars per magnitude interval and peaking at its brighttest limit. The peak of
the VS luminosity function has approximately 25\% more stars than the remaining 
fainter part of the VS, independent of the bin
sizes (see Table 3 and Fig. 7). Therefore, the VS can be described as the
faint tail of a long continuous vertical distribution formed by stars developing
non-degenerate helium cores; the upper part of this long VS is represented by
the so-called ``vertical red clump'' (VRC), recently extensively discussed in the
literature (e.g., Zaritsky \& Lin 1997; Beaulieu \& Sacket 1998; Gallart 1998;
Ibata et al. 1998). The presence of VRC stars in the Hess diagram (density-coded
CMD) of a 2$^o$$\times$1.5$^o$ region located $\sim$ 2$^o$ northwest of the
center of the LMC was interpreted by Zaritsky \& Lin (1997) as red clump stars
that are closer to us than those in the LMC. However, according to Girardi et al.
(1998) and Beaulieu \& Sacket (1998), among others, evolutionary effects 
appear to describe its nature more satisfactorily. Thus, VRC stars 
should be the more massive clump stars, while stars with $M$ $\sim$ 2 
$M_{\odot}$ should define the lower magnitude limit (Girardi's secondary clump); 
stars with even smaller masses are grouped in the RGC. Fig. 4 shows 
the presence of not only VS stars but also VRC and HB stars. Note that even 
though both the Zaritsky \& Lin and our present surveyed areas are roughly similar 
in size, VRC stars are clearly much less numerous than VS stars in our Fig. 4, which 
surprisingly contrasts with their Hess diagrams where no VS stars are seen
despite the presence of the VRC. Certainly, if an LMC field contains 
both high mass (VRC) and low mass (RGC) stars, we should also expect to find 
intermediate-mass stars (VS stars), which were not detected in the Zaritsky \& Lin 
survey data. We are uncertain as to what causes this paradox.

On the other hand, bearing in mind the differences mentioned above, it 
would be interesting to investigate how fundamental properties of VS stars compare 
with those predicted by Girardi for secondary clump stars. Note that Girardi
predicted that secondary clumps should be observed in fields with an
important number of 1 Gyr old stars ($M$ $\sim$ 2 $M_{\odot}$) mixed with older
stars, and with metallicities higher than about $Z$ = 0.004 ([$Fe/H$] $\simeq$ -0.7).
In addition, he pointed out that neither main nor secondary clumps should be
mixed due to differential reddening, distance dispersions and photometric errors.
We first derived the ages of NGC\,2209 and SL\,515 and compared
them with those of Girardi's models. The cluster ages were estimated using 
the $\delta$$T_1$ index as defined in Paper\,I, yielding values of 1.5 and
1.6 Gyr for NGC\,2209 and SL\,515, respectively. These values are in good 
agreement with the ages associated with stars having $M_{Hef}$ just in the limit for
non-degenerate helium cores. We also estimated the  ages for the remaining star
clusters contained in the selected LMC fields, all of them resulting to be on 
average younger than 1.5 Gyr old. The ages derived for SL\,388 and SL\,509 
in Paper\,II are 2.2 and 1.2 Gyr, so that slightly older stars than those predicted 
by Girardi's models could also fall into the CMD VS region. However, most of the 
clusters and surrounding fields of Paper\,II - except ESO\,121-SC03 - have  ages in 
the range 1.0-2.2 Gyr, but only in two of them were secondary clumps clearly 
distinguished. Moreover, the metallicity of SL\,509 is [Fe/H] = -0.85, while those
of the surrounding cluster fields are all in the range $-$0.35 - $-$0.7.

In order to find some explanation for such a paradoxical result, which would appear
to be opposite to the predictions of  Girardi's secondary clump models, we 
counted VS and
RG box stars for all the surrounding fields of clusters analized in Paper\,II, and 
put these values in the VS vs. RG box plane. We applied reddening corrections 
with respect to the SL\,388 and SL\,509  fields ($E(B-V)$=0.03) and adopted the same 
foreground star contamination for all the fields, given the similar galactic star
distribution in their CMDs compared to the field CMD of OHSC\,37 (see Fig. 4 of 
Paper\,II). Fig. 5 shows the resulting relationship (open squares), in which we 
also included the 9 Gyr old field centered on ESO\,121-SC03, the outermost OHSC\,37 
field, and the inner disk SL\,769 field. In particular, the younger SL\,769 fields
turnoffs are younger than 1 Gyr, thus providing an important number of RG stars. 
As can be seen, fields with only a few 
RG box stars do not have many VS stars either, independent of their ages and 
metallicities, while VS stars become more important as the number of RG box stars 
increases. However, the surrounding fields do not seem to show the same
correlation in Fig. 5 as for selected LMC fields (star symbols). In the case of 
SL\,769, the number of VS stars is near the average of those in the selected LMC 
fields, but RG box stars are nearly three times more numerous. Furthermore, the fields
around SL\,388 and SL\,817 share similar ages, metallicities and number of VS stars,
while the number of RG box stars in the SL\,817 field is twice that in the SL\,388
field, which suggests that a large number of RG stars alone is not a sufficient
requirement for the appearance of the VS phenomenon. The number of VS stars in 
the fields of SL\,509 and SL\,862 are also quite different, although their ages, 
metallicities and number of RG box stars are very similar. Furthermore,
SL\,509 itself has 5 VS stars (see Section 3.2), whereas no VS stars appear to be 
associated with SL\,862. All these results apparently suggest that there should be
other conditions, besides age, metallicity and necessary RG star density, 
that would trigger the formation of VS stars, such as the environment of the VS star 
forming regions, different star formation rate, mass function, etc. 
Nevertheless, non-uniform spatial distribution of VS stars in the LMC 
reveals that
non-homogeneously distributed star formation events occurred in this galaxy
about 1-2 Gyrs ago.

\section{CONCLUSIONS} 

>From the analysis of Washington photometry for 21 selected fields located
in the northern part of the LMC, and 14 cluster fields distributed throughout
the LMC disk, we conclusively identify the existence of a vertical structure of
stars that lies below the RGC at its bluest color and up to 0.45 mag fainter.
Our previous data (Paper\,II) uncovered two northern fields which contained
what appeared to be a ``dual clump'', with a secondary clump lying fainter
and bluer than the RGC. Stars lying in the same CMD region were described as
very old stars ($t$ $\sim$ 10 Gyr) by Westerlund et al. (1998) from $BV$ photometry
of three fields located in the NE of the LMC. However, our much larger present
database indicates that there exists a continuous distribution of stars, which
we term VS (``vertical structure'') stars, not only in the CMDs of field 
stars,  but also in certain intermediate-age star clusters. These 
results demonstrate that VS stars belong to the LMC and that they are not 
composed of old objects in the LMC or of a background population of RGC stars. 
We also determine that VS stars are only found in those fields which satisfy some 
particular conditions, such as  containing a significant number of 1-2 Gyr old 
stars and which have metallicities higher than [Fe/H]$\approx$ -0.9 dex, 
in good agreement with Girardi's (1999) models which predicted that a minimum in the 
luminosity of core He burning giants is reached just before degeneracy occurs. These 
conditions constrain 
the VS phenomenon to appear only in some isolated parts of the LMC, particularly
those with a noticeable large giant population. However, a large number of
RG stars, of the appropriate age and metallicity, is not a sufficient requisite 
for forming VS stars. Thus, for example, we found an area spread over 2.6$^{{\Box}^{o}}$ 
centered just to the north-east of SL\,509 with 3 times fewer RG stars than
the inner disk cluster SL\,769, but with approximately the same number of 
VS stars. Clusters with  the appropriate age and metallicity to contain a 
significant number of VS stars are also required to be relatively 
massive; NGC\,2209, for example, constitutes a good example of Girardi's
predictions. Finally, although Girardi's models successfully predict the existence 
of red giants fainter and bluer than RGC stars on the basis of an evolutionary 
effect, there is still a need for more detailed studies explaining for example the 
VS vs. RG relationship, the ratio between the number of VS and VRC stars, whether 
tilted RGCs are related with VS features, the VS luminosity function, etc. The 
fact that Zaritsky \& Lin (1997) found red clump stars with high and low masses, but
none at the intermediate degenerate mass limit to form VS stars also remains 
unexplained. Indeed, 1-2 Gyr old stars with [Fe/H] $\sim$ -0.3 dex are very common 
in the LMC, although VS stars are only clearly seen in certain parts of the galaxy, which
constitutes an unresolved mystery.

The authors would like to thank the CTIO staff for their kind hospitality
during the observing run. A.E.P. greatly appreciates the opportunity provided
by CTIO of spending two months of his Gemini Fellowship at its Headquarters
in La Serena, Chile. Support for this work was provided by the National Science 
Foundation through grant number GF-1003-98  from the Association of Universities 
for Research in Astronomy, Inc., under NSF Cooperative Agreement No. 
AST-8947990. Peter Stetson is also sincerely acknowledged for his help
in the installation and execution of DAOPHOT and DAOMASTER. D.G. would like to
acknowledge a very useful conversation with D. Alves who pointed out the
importance of observing NGC\,2209. We also thanks the referee for his 
valuable comments and suggestions. A.E.P. and J.J.C. acknowledge the
Argentinian institutions CONICET and CONICOR for their partial support.
J.F.C.S. Jr. also  acknowledges the Brazilian institutions CNPq and FAPEMIG for their 
support.

\newpage

\label{fig1}
\begin{figure}
\figcaption{Schematic mosaic of the 21 fields observed in the north of the LMC.
The number of VS stars in each field is also indicated. Note the strong concentration 
just north-east of the cluster SL\,509. Fields with an asterisk have tilted RGC 
(see Section 3.1 for details).}
\end{figure}

\begin{figure}
\label{fig2}
\figcaption{Typical Washington $T_1$ vs. $C-T_1$ CMD for a selected field in the 
northern part of the LMC. Each field contains on average 12,000 stars.}
\end{figure}

\begin{figure}
\label{fig3}
\figcaption{Enlargement of the RGC region in the Washington $T_1$ vs. $C-T_1$ CMD 
for each selected LMC field. Field identification is also shown (see Table 1).}
\end{figure}

\begin{figure}
\label{fig4}
\figcaption{Composite Washington $T_1$ vs. $C-T_1$ CMD using all the measured stars
in the selected LMC fields. The SL\,388 and SL\,509 cluster fields are also included 
(see Section 3.2 for details). Note that VRC and HB stars are also present 
(see Section 3.4 for details).}
\end{figure}

\begin{figure}
\label{fig5}
\figcaption{Relationship between the number of field VS stars and the number of
RG box stars. Symbols represent selected LMC fields 
($\star$), NGC\,2209 field ($\triangle$), and previously published LMC cluster 
fields ($\Box$) (see Sections 3.2 and 3.4 for details).}
\end{figure}

\begin{figure}
\label{fig6}
\figcaption{Washington $T_1$ vs. $C-T_1$ CMD for the NGC\,2209 field, which is 
located $\sim$ 14$^o$ away from the selected fields in the northern part of the 
LMC. A line following the direction of the reddening vector ($\Delta$$E(B-V)$ = 0.20) 
is also shown.}
\end{figure}

\begin{figure}
\label{fig7}
\figcaption{Luminosity function for VS stars.}
\end{figure}
\end{document}